\documentclass[traditabstract]{aa} % for a referee version ,referee
\usepackage{txfonts}
\usepackage{graphicx}

\begin{document}

\title{Gravity-Selected Galaxy Clusters: a Tight Mass-Richness relation and an unclear Compton $Y$-richness trend.}
\titlerunning{Gravity-Selected Clusters: Tight Scaling Relations with Puzzling Outliers}
\author{S. Andreon$^{1}$\thanks{E-mail:stefano.andreon@inaf.it}, M. Radovich$^{2}$}
\authorrunning{Andreon \& Radovich}
\institute{
$^1$ INAF--Osservatorio Astronomico di Brera, via Brera 28, 20121, Milano, Italy\\
$^2$ INAF--Osservatorio Astronomico di Padova, Vicolo Osservatorio 5, 35122, Padova, Italy\\
}
\date{Accepted ... Received ..; in original form ..}
\abstract{
This paper, the third in a series, investigates the scaling relations between optical richness, weak-lensing mass, and Compton $Y$ for a sample of galaxy clusters selected purely by the effect of their gravitational potential on the shapes of background galaxies. This selection method is uncommon, as most cluster samples in the literature are selected based on signals originating from cluster baryons. We analize a complete sample of 13 gravity-selected clusters at intermediate redshifts (with $0.12 \leq z_{phot} \leq 0.40$) with weak-lensing signal-to-noise ratios exceeding 7.
We measured cluster richness by counting red-sequence galaxies, identifying two cases of line-of-sight projections in the process, later confirmed by spectroscopic data. Both clusters
are sufficiently separated in redshift that 
contamination in richness can be straighforwardly dealt because the two red sequences do not blend each other. 
We find an exceptionally tight richness--mass relation using our red-sequence-based richness estimator, with a scatter of just $\sim0.05$ dex, 
smaller than the intrinsic scatter of Compton Y with mass for the same sample. The lower scatter highlights the effectiveness of 
richness 
compared to 
Compton $Y$.
No outliers are found in the richness-mass scaling, whether or not one cluster with a mass likely affected by projection effects is included in the sample.
In the Compton $Y$-richness plane, the data do not delineate a clear trend. The limited sample size is not the sole reason for the unclear relation between Compton $Y$ and richness, since the same sample, with identical richness values, exhibits a highly significant and tight mass-richness correlation.
}
\maketitle

\section{Introduction}

The mass of galaxy clusters is a fundamental quantity in both astrophysics and cosmology. Precise mass estimates are crucial for a range of applications, including studies of the physical processes shaping the intracluster medium (ICM) and galaxy populations, cosmological tests using the cluster mass function, and the comparison of clusters with different properties, since more massive clusters tend to host more galaxies, gas, and dark matter, comparisons must be properly scaled by mass.
However, cluster mass is not directly observable. The two primary methods that directly measure mass, weak gravitational lensing and the caustic technique, are observationally expensive: weak lensing (Tyson et al. 1990; Kaiser \& Squires 1993; Schneider 2006) requires precise measurement of small distortions in the shapes of faint, distant galaxies, while the caustic method (e.g., Diaferio \& Geller 1997) demands dense spectroscopic sampling to trace escape velocities.

Forthcoming wide-field optical and near-infrared surveys, such as \textit{Euclid} (Laureijs et al. 2011), \textit{Roman} (Green et al. 2012), Rubin Observatory LSST (LSST Science collaboration 2009), and the China Space Station Telescope (CSST, Shan 2011), will detect millions of clusters over large areas of the extragalactic sky. However, only a small fraction will have direct mass estimates, and those that do will predominantly lie at lower redshifts. For instance, \textit{Euclid} is expected to detect nearly two million clusters peaking at $z \sim 0.8$, with a tail extending to $z=2$ (Sartoris et al. 2016), but only about 30 thousands (1.5\%) will have a weak-lensing signal-to-noise above 5, and these will be limited to $z<0.7$, with mostly at $0.2<z<0.4$  (Andreon \& Berg\'e 2010). 

Given the lack of direct mass measurements for the bulk of cluster samples in optical and near-infrared surveys, robust and efficient mass proxies that can be derived from this type of data are essential. Galaxy cluster richness, particularly when restricted to red-sequence galaxies that are largely devoid of gas, has proven to be a reliable and observationally inexpensive mass proxy. Its robustness stems from the luminosity and color of red galaxies being minimally affected, if any, by interactions, cluster mergers and, in general by the cluster dynamical state. Moreover, photometric redshifts and red-sequence colors (e.g., Gladders \& Yee 2000) enable the detection of projected structures along the line of sight, provided the redshift separation exceeds $\Delta z = 0.02$ (Andreon \& Berg\'e 2010).

Most studies adopting a richness--mass relation assume a linear relation in log quantities without outliers (e.g., Saro et al. 2015; Jimeno et al. 2018; Costanzi et al. 2019; Bellagamba et al. 2019, Bleem et al. 2020; Singh et al. 2024; Ghirardini et al. 2024). Empirically, very few outliers are known: none among 53 clusters in Andreon \& Hurn (2010), one of 11 in Fo{\"e}x et al. (2012), none of 23 in Andreon \& Condon (2014), and only two dubious cases in samples of 39 and $\sim$20 clusters in Andreon (2015) and Mantz et al. (2016), respectively. Andreon et al. (2024, hereafter Paper I) reported one outlier in a sample of four. Notably, all but Paper I rely on X-ray-selected samples, whereas future surveys like \textit{Euclid}, \textit{Roman}, Rubin, and CSST will select clusters using galaxy overdensities. X-ray selection introduces biases (e.g., Vikhlinin et al. 2009,  Andreon et al. 2011, 2016, 2017a), 

It is therefore valuable to characterize the richness--mass scaling relation in a sample that is not selected via the ICM. This is the first aim of our paper, where we analyze the gravity-selected sample of Andreon \& Radovich (2025, hereafter Paper II), using the very same richness definition of one of those adopted in the \textit{Euclid} cluster pipeline (Euclid collaboration: Mellier et al. 2024). 

A second aim of this paper is to investigate the SZ signal--richness scaling relation, where SZ signal is often loosely referred as mass. Current findings are inconclusive, due to the limited data and heterogeneous definitions of richness and SZ observables. Gonzalez et al. (2019) identified outliers in the SZ-richness relation, two of which are known mergers with suppressed Compton $Y$ values. Di Mascolo et al. (2020) expanded the sample and argued that these outliers may simply reflect statistical scatter in a poorly sampled relation, and further proposed that two of their own clusters might be genuine outliers. Their analysis is complicated by filtering in their SZ maps, which forces extrapolation of Compton $Y$ to poorly sampled spatial scales. Orlowski-Scherer et al. (2021) also reported evidence of two populations in this plane, but, as detailed in Paper II, it to be verified wether their systems without an SZ signal are genuine clusters or line-of-sight projections misidentified as single clusters. 
Again, some studies (e.g. Grandis et al. 2021, Bleem et al. 2020) assume a gaussian scatter of the SZ observable and richness with mass, and as consequence a gaussian scatter between the SZ observable and richness.

Throughout this paper, we use the word baryon in the astronomical sense, including electrons, responsible for both thermal bremstralung emission and SZ effect. We assume $\Omega_M=0.3$, $\Omega_\Lambda=0.7$, and $H_0=70$ km s$^{-1}$ Mpc$^{-1}$. 
Gaussian posteriors are summarized
in the form $x\pm y$, where $x$ and $y$ are 
the mean and standard deviation. 
Non Gaussian posteriors are summarized as $x^{+y}_{-z}$ where $(x-z),x$, and $x+y$ are the
$(16,50,84)$ percentiles. All logarithms are in base 10.

\begin{figure*}
\centerline{%
\includegraphics[trim=40 20 100 70,clip,width=7truecm]{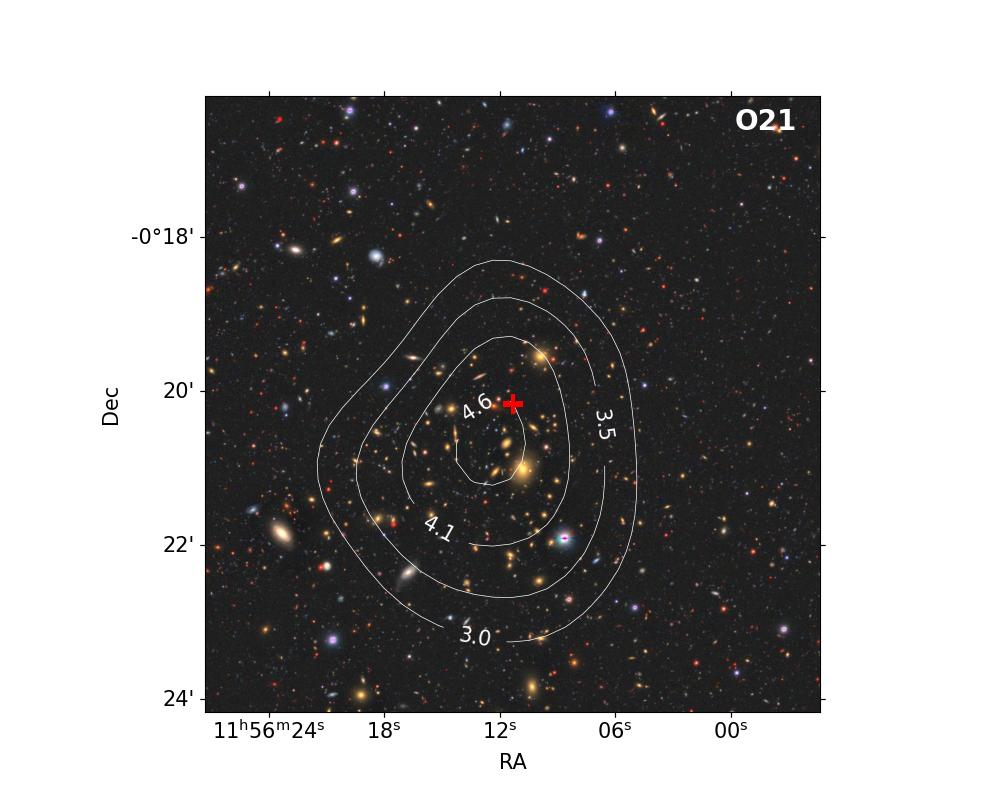}
\includegraphics[trim=40 20 100 70,clip,width=7truecm]{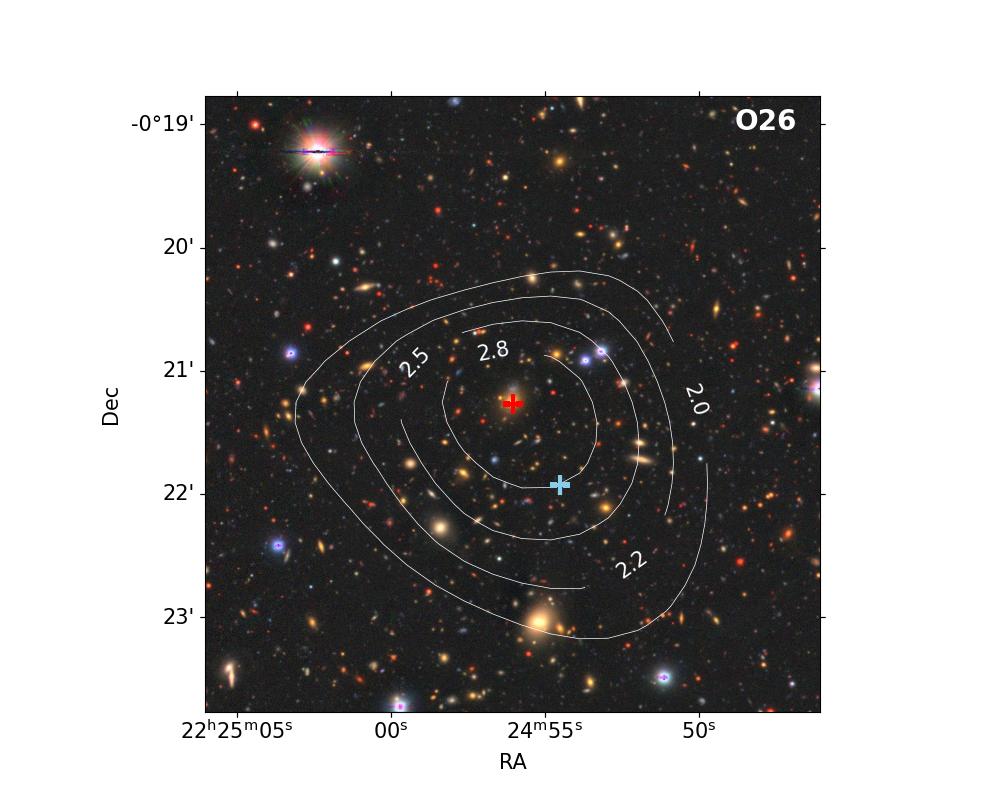}}
\centerline{%
\includegraphics[trim=40 20 100 70,clip,width=7truecm]{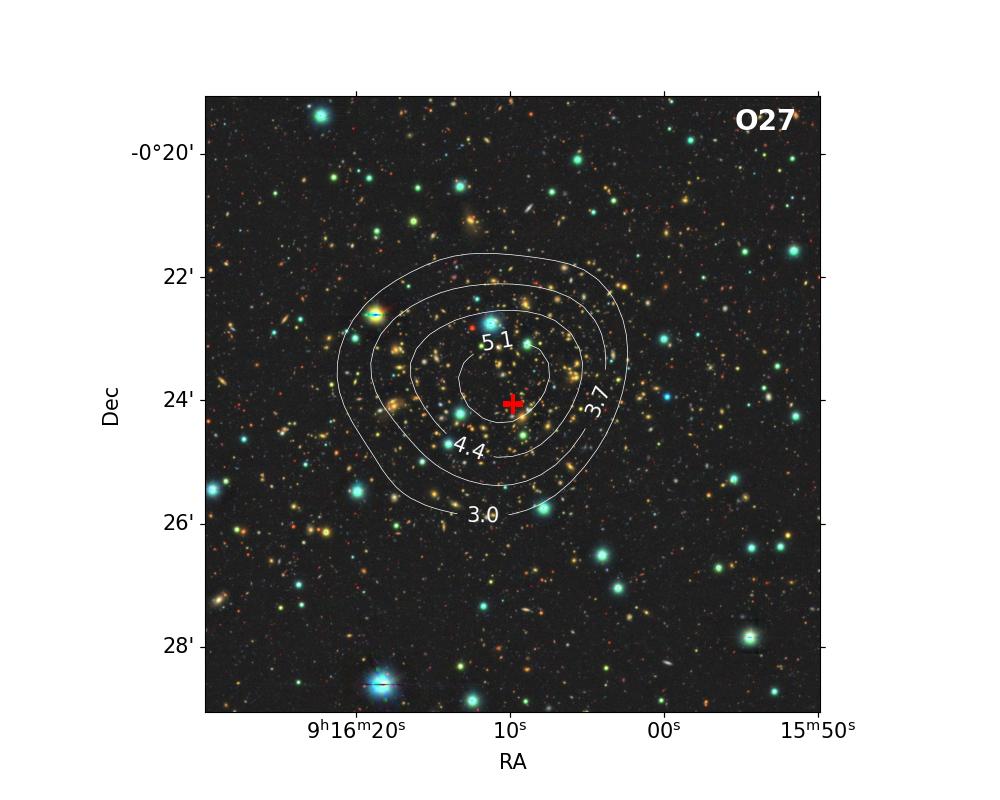}%
\includegraphics[trim=40 20 100 70,clip,width=7truecm]{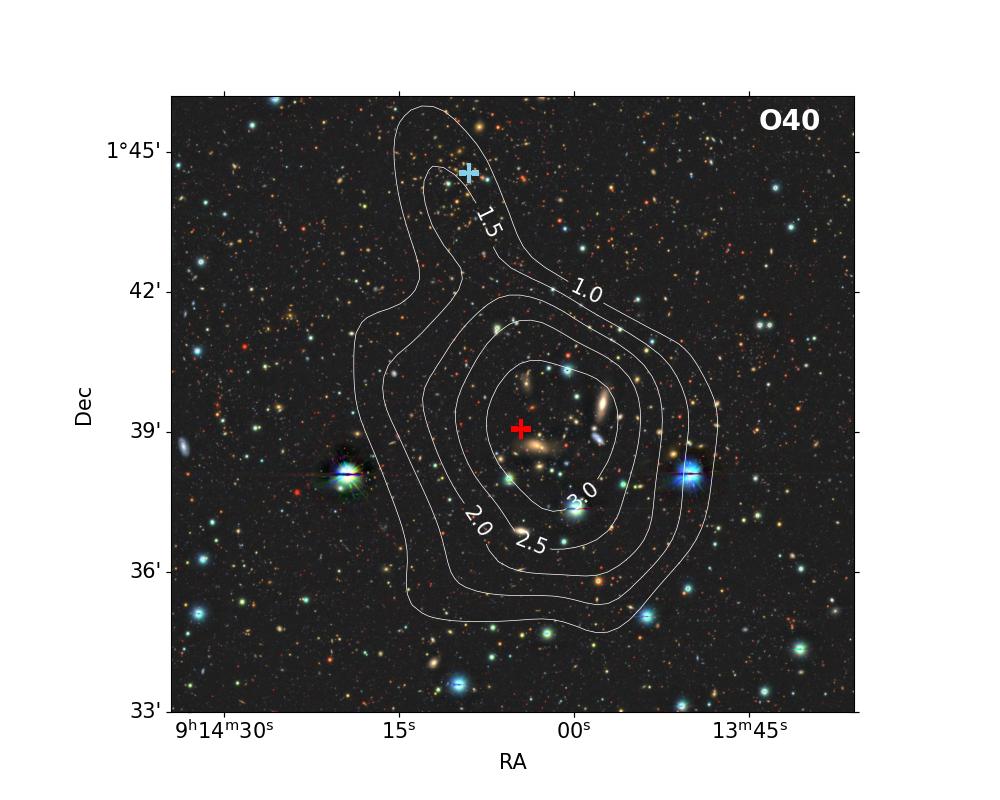}}%
\caption[h]{True-color ($grz$) HSC images with overlaid contours of the weak-lensing S-map, which serves as a proxy for signal-to-noise, convolved with a Gaussian filter of $\sigma = 2$ arcmin for display purposes. The  red and  cyan crosses mark the positions of the target cluster and other clusters (present in O26 and O40 panels), respectively. The true-color images are taken from https://www.legacysurvey.org/viewer}
\label{fig:maps}
\end{figure*}

\setlength{\tabcolsep}{6pt}

\begin{table*}
\caption{Cluster sample and summary of analysis results.
}
\begin{tabular}{lrrrrrrrrrrrrrr}
\hline\hline
ID & \multicolumn{1}{c}{R.A.} & Dec &  z$_{\mathrm spec}$ & logM$_{200}$ & err & logY$_{200}$ & Y err & $n_{200,WL}$ & err & R.A. & Dec & $\Delta r$ & $n_{200,A16}$ & err \\
 & \multicolumn{2}{c}{J2000} & & [M$_\odot$] & dex & Mpc$^{2}$ & 10$^{-6}$ Mpc$^{2}$ & & &  \multicolumn{2}{c}{J2000} & [Mpc] & & \\
\hline
 (1) & (2) & (3) &  (4) & (5) & (6) & (7) & (8) & (9) & (10) & (11) & (12) & (13) &  (14) & (15) \\
\hline
O3 &  37.9224 & -4.8789 & 0.185   & 14.81 & 0.09    & -4.33 & 10.4  & 80 & 10  & 37.9218 & -4.8792 & 0.0 & 80 & 10        \\
O7 &  336.0368 & 0.3293 & 0.146   & 14.68 & 0.11    & -5.08 & 5.6  & 33 & 8  & 336.0484 & 0.3298 & 0.1 & 31 & 7 	 \\
O12 &  180.4305 & -0.1903 & 0.167   & 14.53 & 0.14    & -4.64 & 5.3  & 56 & 9  & 180.4354 & -0.1890 & 0.1 & 45 & 8 	 \\
O15 &  177.5865 & -0.6026 & 0.137   & 14.43 & 0.14    & -5.07 & 4.8  & 37 & 7  & 177.5781 & -0.6010 & 0.1 & 32 & 7 	 \\
O19 &  130.5909 & 1.6474 & 0.422   & 15.06 & 0.10    & -5.05 & 13.4  & 107 & 14  & 130.5845 & 1.6467 & 0.1 & 106 & 13 	 \\
O21 &  179.0472 & -0.3361 & 0.256   & 14.78 & 0.11    & -4.42 & 8.1  & 60 & 10  & 179.0479 & -0.3584 & 0.3 & 44 & 8	  \\
O22 &  139.7018 & 2.2062 & 0.284   & 14.68 & 0.19    & -4.77 & 8.2  & 65 & 10  & 139.6948 & 2.2067 & 0.1 & 61 & 9	  \\
O27 &  139.0410 & -0.4010 & 0.329   & 15.04 & 0.11    & -4.25 & 11.9  & 127 & 14  & 139.0416 & -0.3920 & 0.2 & 125 & 13   \\
O28 &  223.0843 & -0.9708 & 0.313   & 14.67 & 0.14    & -4.51 & 10.2  & 33 & 8  & 223.0843 & -0.9708 & 0.0 & 30 & 7	  \\
O32 &  217.6835 & 0.8127 & 0.321   & 14.74 & 0.14    & -5.04 & 12.3  & 63 & 9  & 217.6727 & 0.8058 & 0.2 & 56 & 9 	 \\
O40 &  138.5191 & 1.6512 & 0.166   & 14.51 & 0.16    & -4.96 & -  & 20 & 6  & 138.4900 & 1.6543 & 0.3 & 17 & 5 	 \\
O48 &  336.4214 & 1.0749 & 0.296   & 14.50 & 0.17    & -4.94 & 8.0  & 41 & 9  & 336.4215 & 1.0754 & 0.0 & 43 & 9 	 \\
\hline
O26 &  336.2335 & -0.3545 & 0.315   & 14.46 & 0.21    & -5.21 & 6.4  & 27 & 7  & 336.2258 & -0.3478 & 0.2 & 26 & 6	  \\
\hline
\end{tabular}
\hfill \break 
{\bf Notes.}
The table provides the following information: cluster ID (Col. 1); coordinates (RA and Dec, Cols. 2 and 3); spectroscopic redshift z$_{\mathrm spec}$ (Col. 4); weak-lensing mass  logM$_{200}$ and its uncertainty (Cols. 5 and 6); spherically integrated Compton-Y parameter  logY$_{200}$ and its uncertainty (Cols. 7 and 8); richness $n_{200,WL}$ within the weak-lensing 
radius, centered on the coordinates given in Cols. 2 and 3 (Cols. 9 and 10); coordinates of the barycenter of the galaxy population (RA and Dec, Cols. 11 and 12); positional offset between the barycenter and the coordinates in Cols. 2 and 3 (Col. 13); and richness $n_{200,A16}$ within the radius inferred from galaxy counts, along with its uncertainty (Cols. 14 and 15).
Except for O26, for which all quantities have been derived in this paper, coordinates in Columns 2 and 3 are from Oguri et al. (2021) and Colums 4 to 8 are from Paper II. O26 weak-lensing mass is affected by projection effects. Errors in richness are approximated as Gaussian in this Table. Numbers needed to derive the exact distribution of richness errors, used in our fit, are listed in Table~\ref{tab2}.
The Compton Y value reported for O40 is the 95\% upper limit. \hfill
\label{tab1}
\end{table*}

\begin{figure}
\centerline{\includegraphics[width=8truecm]{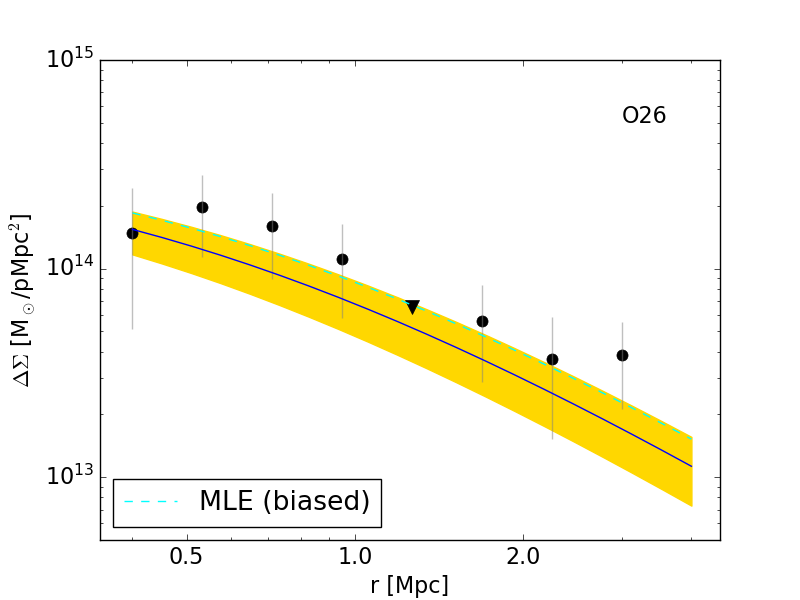}}
\caption{Binned tangential shear profile of the O26 cluster, incorporating the updated identification. The solid line, with yellow shading, represents the mean model and the 68\% uncertainty region. The uncertainty in the model also includes intrinsic scatter, while the error bars plotted account only for shape noise and large-scale structure. The dashed cyan line represents the maximum likelihood estimate (MLE), which is biased. The triangle denotes the $2\sigma$ upper limit.
}
\label{fig:WL}
\end{figure}

\begin{figure}
\centerline{\includegraphics[width=8truecm]{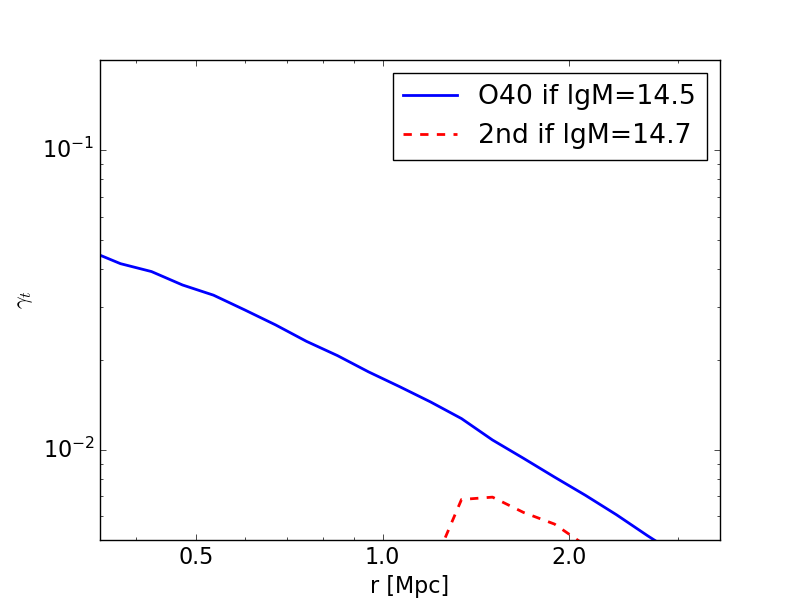}}%
\caption[h]{ Expected  O40 tangential shear profile if O40 had the fitted mass value ($\log M/M_\odot=14.5$), and of a $\log M/M_\odot=14.7$ cluster located 5.5 arcmin from O40. The figure demonstrates the negligible contribution of the contaminating cluster. 
}
\label{fig:contam}
\end{figure}

\begin{figure}
\centerline{\includegraphics[width=8truecm]{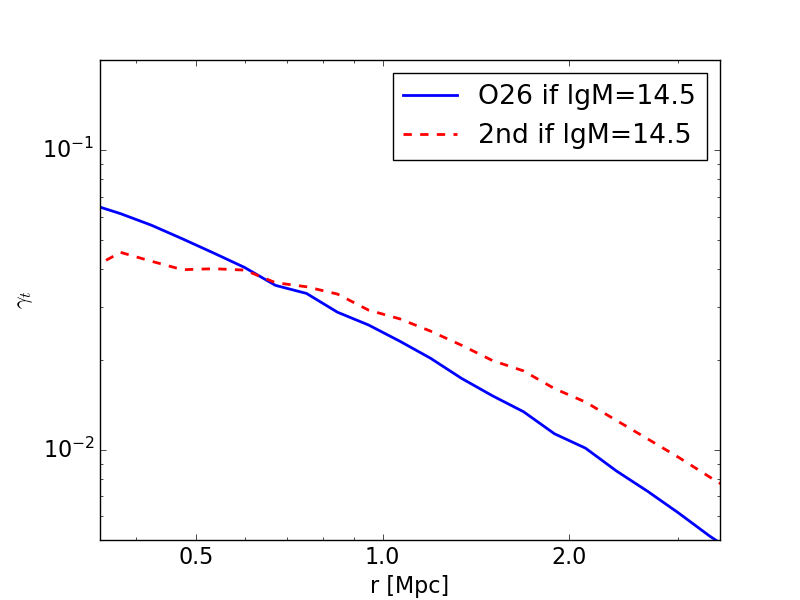}}%
\caption[h]{ Expected  O26 tangential shear profile if O26 had the fitted mass value ($\log M/M_\odot=14.5$), and of a $\log M/M_\odot=14.5$ cluster located 100 arcsec from it. The figure demonstrates the possible large contribution of the contaminating cluster. 
}
\label{fig:contamO26}
\end{figure}

\begin{figure}
\centerline{\includegraphics[width=8truecm]{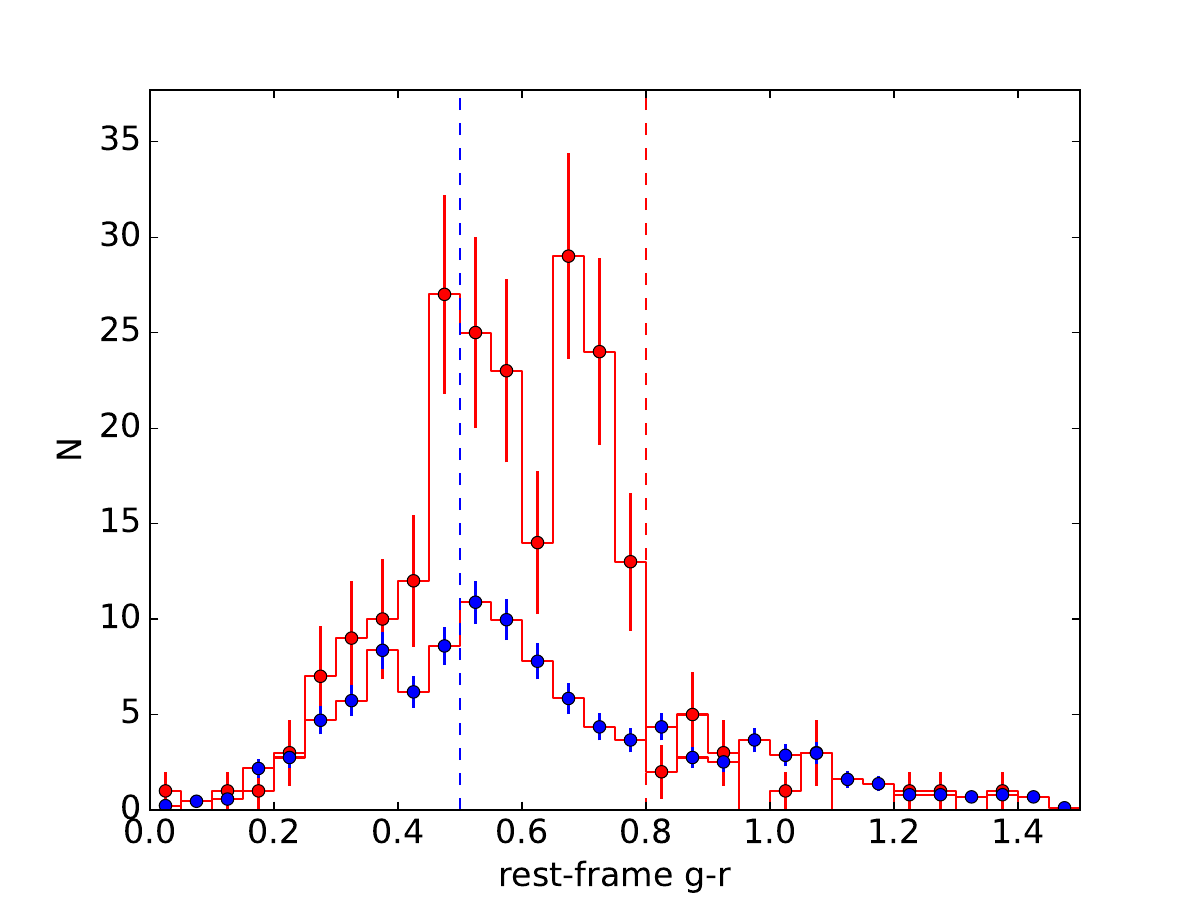}}%
\caption[h]{Color distribution of galaxies within $r_{200,\rm wl}$ (red histogram) and in a control area (normalized to the cluster solid angle, blue histogram) for O32. The vertical lines indicate the expected color range of the red sequence. The peak on the right corresponds to the color expected for the cluster redshift, while the second peak represents a contaminating structure along the line of sight spectroscopically confirmed at $z=0.136$ (see text for details). For O32, the color range was reduced by 0.1 mag on the blue side to exclude this contamination. 
Only galaxies brighter than $M^e_V=-20$ mag, assuming all galaxies are at the O32 redshift, are considered, which boost the low redshift contamination.
}
\label{fig:CMO32}
\end{figure}

\section{Sample, Measurements, and Results}

\subsection{Sample}

The sample analyzed in this study is identical to that used in Paper II. It is a purely gravity-selected sample, constructed from the shear peak list produced by Oguri et al.\ (2021), from which we selected a complete sample formed by the 13 clusters within the photometric redshift in the range $0.12 \leq z \leq 0.4$ and a weak-lensing signal-to-noise ratio greater than 7. This high threshold ensures that the Eddington (1913) correction and its associated uncertainty are negligible. Spectroscopic redshifts are derived in Paper II from SDSS DR18 spectroscopic data (Almeida et al.\ 2023). Paper II also derives the spherical Compton-$Y$ parameter within $r_{200}$ using the ACT Compton-$y$ map (Coulton et al.\ 2023).

The analysis presented in next sections of our paper show that the weak-lensing mass of O26 is likely affected by projection effects, and therefore this object should be treated with caution. For safety reasons, we decided to drop it from the reference sample, although we plot O26 in all plots and we re-inserted it in our sample to test the (negligible) effect of ignoring or including this object. When O26 is excluded, it is excluded because of its likely affected weak-lensing mass, whereas richness measurement is possible and performed.
Since the object exclusion is unrelated to richness, the sample without O26 continues to be gravity-selected and our analysis does not need to account for a never-applied richness selection.

\subsection{Weak-lensing masses}

Cluster masses are estimated in Paper II by fitting the tangential shear profile, derived from the HSC DR1 shape catalog (Mandelbaum et al.\ 2018), using a Navarro, Frenk, \& White (1997) model over a radial range of $0.35$ to $3.5$ Mpc. This range is selected to mitigate the effects of dilution, mis-centering, deviations from the weak-lensing regime, and contamination from neighboring clusters and large-scale structures. 
The fitting procedure incorporates the concentration--mass relation of Dutton \& Macci\`o (2014), assuming uniform priors on the logarithm of the concentration, and uses the Tinker et al.\ (2008) mass function to account for Eddington (1913) bias. An intrinsic scatter of 20\% in the lensing signal is included to account for cluster elongation, triaxiality, and correlated halos. Further methodological details are provided in Papers I and II.

Figure~\ref{fig:maps} presents the weak-lensing signal-to-noise maps for the four clusters individually discussed in our paper, based on the Schirmer et al.\ (2004) S-statistics and convolved with a Gaussian filter of $\sigma = 2$ arcmin for display purposes only (see Radovich et al.\ 2008 for details). 
O26 has primary identification in Oguri et al.\ (2021) with the cluster at ra,dec=(336.2271,-0.3654) and $z_{phot}=0.141$ that using
SDSS spectroscopic data we confirmed in Paper II to be at the spectroscopic redshift of $z = 0.142$. This cluster is about 100 arcsec South
of the weak-lensing peak shown in Fig.\ref{fig:maps}, which in turn is coaligned with 
X-ray emission (measured from a short Chandra exposure targeting a different object, ObsID 3962) of a different cluster 
with spectroscopic redshift (of the brightest cluster galaxy and a few other galaxies)
of $z = 0.315$. We now adopt as primary identification the latter cluster, that was previously  the secondary identification in Oguri et al.\ (2021). %at coordinates and redshift different from the primary identification in Oguri et al.\ (2021). Specifically, the brightest cluster galaxy (BCG) of O26 is located at (RA, Dec) = (336.2335, $-0.3545$) with redshift $z = 0.315$, approximately 100 arcseconds north of the BCG of the originally identified cluster in Oguri et al.\ (2021), which has $z = 0.142$. 
Figure~\ref{fig:WL} shows the revised tangential shear profile for cluster O26 after center and redshift change. The inferred mass changes by approximately $1\sigma$, from $\log M/M_\odot = 14.22 \pm 0.22$ to $\log M/M_\odot = 14.45 \pm 0.21$. However, the shear signal is likely contaminated by the cluster initially identified as O26 due to the low angular separation and similar mass, as discussed later.

About O40,
approximately 5.6 arcmin north of it (corresponding to about 1 Mpc at the cluster redshift), there is a known cluster, RM J091409.2+014405.3 (Rozo et al.\ 2015; Radovich et al.\ 2017). Given its limited impact on the shear map, particularly after projecting the shear tangentially to O40, we do not expect it to significantly contaminate the tangential shear profile of O40. Nonetheless, we investigate this possibility.
Figure~\ref{fig:contam} compares the expected tangential shear profile of a cluster with $\log M/M_\odot = 14.5$ (i.e., the mass measured for O40) along with the contribution from a background cluster at a projected separation of 1 Mpc, assumed to have $\log M/M_\odot = 14.7$ at $z = 0.389$. The assumed redshift correspond to the spectroscopically redshift of RM J091409.2+014405.3 determined using SDSS DR18 data (Almeida et al.\ 2023) as we did for the targets, whereas
the estimated mass correspond to the measured richness-based mass of the contaminant, derived as described below for the targets.
The contaminating cluster contributes negligibly to the total shear, and only at radii greater than 1 Mpc, as expected. This contamination can, therefore, be safely neglected.

\subsection{Richness measurements}

Cluster richnesses are derived using photometry from the third HSC data release (Aihara et al.\ 2022), following the methodology of Paper I, which itself is based on Andreon (2015, 2016) with minor updates and is one of the two richness estimates adopted by Euclid. Briefly, galaxies located on the red sequence and brighter than the passively evolved threshold of $M^e_V = -20$ mag are counted within a radius $r_{200}$. Our operational definition of ``on the red sequence" includes galaxies within 0.1 mag redward and 0.2 mag blueward of the color--magnitude relation, with two exceptions discussed below. The color of the red sequence is self-calibrated from the data, as described in Paper I. We use Kron magnitudes as proxies for total fluxes and employ 1.1 arcsec aperture photometry, measured on images PSF-matched to 1.1 arcsec, for color estimation. The contribution of background and foreground galaxies is estimated using an annulus between 3 and 7 Mpc, with corrections for contamination by unrelated structures. To minimize bias, the annulus is divided into octants, and the two with the highest and the two with the lowest galaxy counts are excluded from the background estimate. 

We compute two richness estimates: (i) the reference richness, $n_{200,WL}$, measured within the weak-lensing radius $r_{200,WL}$ and centered on the Oguri et al.\ (2021) cluster coordinates (as revised by us for O26); and (ii) an additional richness, $n_{200,A16}$, derived iteratively using a radius--richness scaling relation calibrated on clusters with known masses (Andreon 2015), including evolutionary corrections. The aperture for the latter is centered on the barycenter of the galaxy distribution, iteratively determined from the spatial distribution of galaxies within a 1.0 Mpc radius, as described in Andreon (2016) and Paper I. While $n_{200,WL}$ shares the radius and center with the weak-lensing mass measurement, $n_{200,A16}$ does not.

The richness determination identifies two cases of contamination of other structures along the line of sight via the presence of bimodal color distributions.
The redder peak of peak of O32 and O26 color distributions corresponds to the expected color of the red sequence at the target redshift (see Fig.~\ref{fig:CMO32} for O32).
About O32, spectroscopy (DESI Collaboration, 2025) confirms that the redder peak originates from galaxies at the cluster redshift, while the bluer peak (approximately 0.2 mag bluer) comprises galaxies unrelated to O32: among the 21 galaxies with a spectroscopic redshift in this secondary peak, none are at the cluster redshift; many are at $z = 0.136$. Since the contamination is at lower redshift of the target cluster, it is sampled down by about 2.2 magnitudes fainter than the $M^e_V = -20$ mag threshold adopted for O32, leading to a large peak in the color distribution (see Fig.~\ref{fig:CMO32}).
Additionally, the adopted filters are appropriate for the O32 cluster but sample redward of the Balmer break  for the foreground population, making narrow the color distribution of the contaminants because
galaxies with different stellar ages have similar colors in band pairs redder than the Balmer break (Fukugita 1995; Frei \& Gunn 1994). 
As for the target,
we measure the richness of the O32 cluster, and its richness-based mass following Andreon (2016), as updated in Sec.~2.5. We found it to be a $<1$ 10$^{14}$
M$_\odot$, smaller than the error on O32 mass and therefore this contamination turns out to be  negligible for the weak-lensing mass estimation. About
the richness, to remove the contamination is sufficient to adopt a stricter color selection: galaxies must lie no more than 0.1 mag blueward of the red sequence (instead of the default 0.2 mag).

O26 is contaminated by the foreground optical cluster originally identified as O26 in Oguri et al. (2021). This results in a similarly bimodal color distribution. Also in this case we used the stricter color selection to measure the richness. However, the richness-based mass of both the target and the contaminant turned out to be pretty similar, both having a richness-based mass of about $\log M/M_\odot = 14.5$, making the contamination of the foreground not negligible in the weak-lensing estimation (see Fig.~\ref{fig:contamO26}). Therefore, in our reference analysis we removed O26 from our sample of clusters used for scaling relation, reinserting it back to check the sensitivity of our conclusions to the inclusion of this object. As mentioned, dropping this object does not alter the nature of gravity-selected of our sample.
We verified that applying this stricter color selection to the rest of the cluster sample does not alter their measured richness values.

\begin{figure}
\centerline{\includegraphics[width=9truecm]{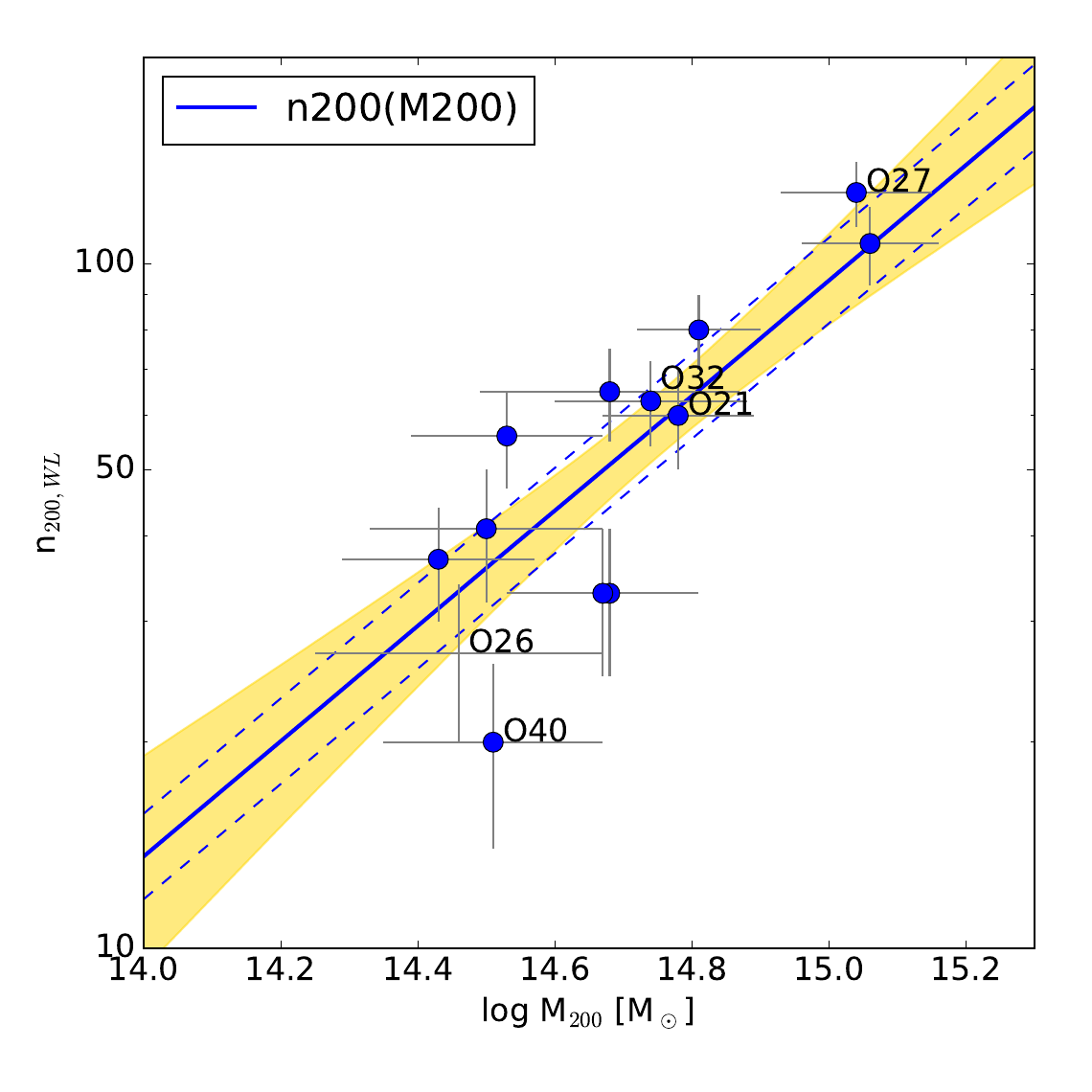}}
\caption[h]{Richness vs. mass scaling relation within $r_{200,WL}$. The solid line with yellow shading represents the mean richness-mass model with 68\% uncertainty, while the dashed corridor indicates the mean model $\pm1$ intrinsic scatter. Clusters discussed in the text are labeled. O26 is not fitted because of its likely contaminated weak-lensing mass. Including it in the sample would not alter the scaling relation. In this plot and in Fig.~\ref{fig:MrichA16}, richness errors are approximated as Gaussian for 
plotting purposes, although their exact distributions are properly accounted for in the fit (see text for details).} 
\label{fig:Mrich}
\end{figure}

The derived richness values and the cluster barycenters are listed in Table~\ref{tab1}. The mean (log) richness error is 0.07 dex.
 The two richness estimates (with fixed or to-be-determined center and radii) are in strong agreement, differing on average by $6 \pm 10$\%, while their mean statistical uncertainty is 19\%. The cluster centers reported by Oguri et al.\ (2021) and the galaxy barycenters computed in this work are also highly consistent: in 9 out of 13 cases, the positional offset is $\leq 0.1$ Mpc (see Table~\ref{tab1}). In the remaining cases, the offsets reflect differences in center definitions. For instance, the offset observed for O27 arises from the lack of spherical symmetry in this cluster, which, as discussed below, is undergoing a merger. Even in cases where the centers differ, the richness estimates remain consistent to within $<1\sigma$. This robustness is due to the fact that the positional offsets are small compared to the aperture size used for richness measurement, and the radial richness profile has a shallow slope near $r_{200}$, making richness relatively insensitive to small shifts in center position.

To assess the covariance introduced by using a common measurement radius for both the mass and the $n_{200,WL}$ richness, we performed measurements at different radii in steps of $\sqrt[3]{1.2} \, r_{200,WL}$. The results indicate that the covariance induced by the shared radius is subdominant compared to the richness measurement uncertainty. Consequently, this source of covariance is not included in the analysis. In Section~\ref{Mn}, we verify whether the conclusions of this study are affected when adopting the iteratively computed richness, which lacks radial covariance with the weak-lensing mass due to its independently determined radius and center.

\subsection{Results using the weak-lensing aperture $r_{200,WL}$}

\begin{figure}
\centerline{\includegraphics[width=9truecm]{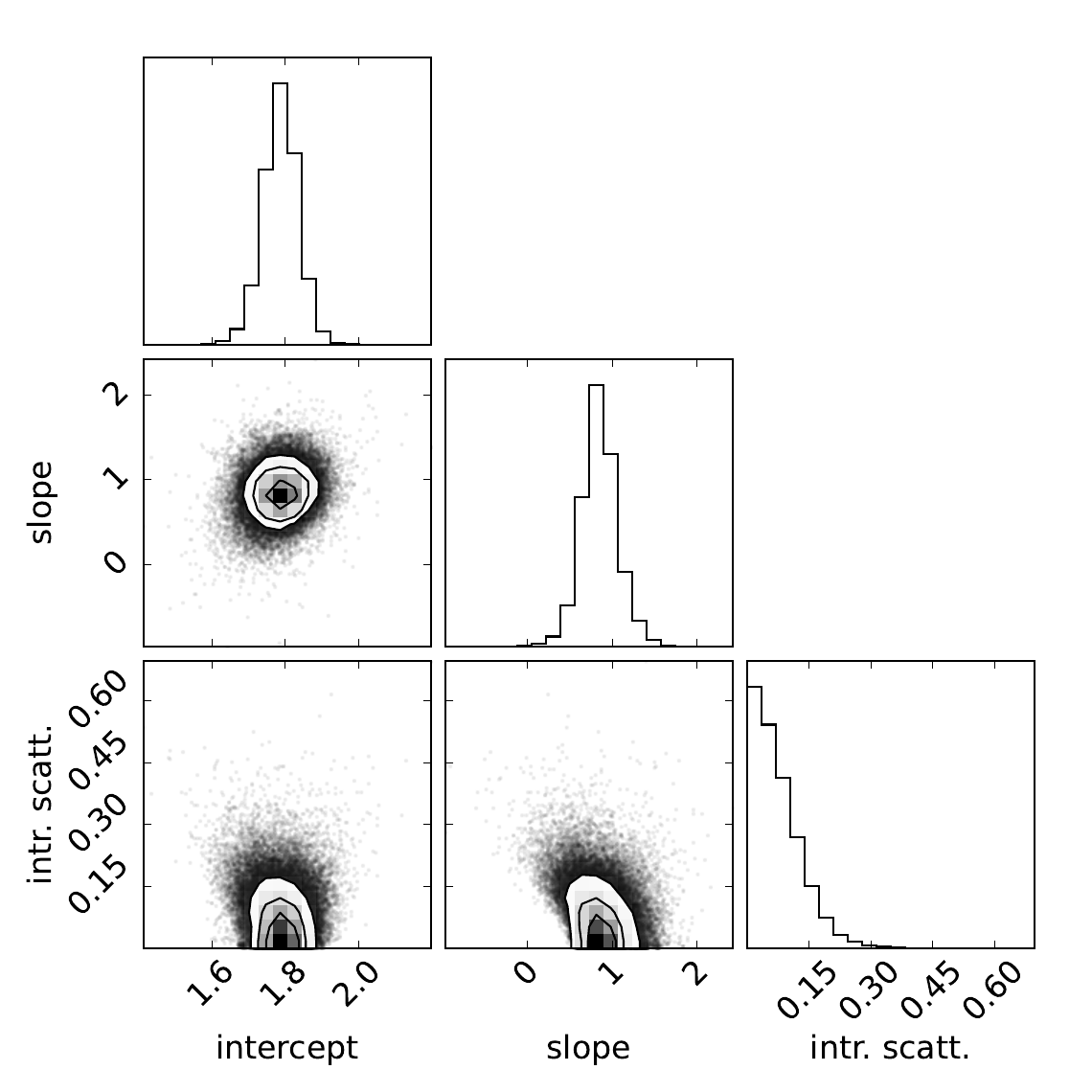}}
\caption[h]{Marginal  (on the diagonal) and joint  
(remaining panels) probability distributions of the parameters in eq.~1.}
\label{fig:corner}
\end{figure}

Figure~\ref{fig:Mrich} shows the relation between mass and richness $n_{200,WL}$ measured within the weak-lensing aperture. We fit the sample with a linear relation (between logarithmic quantities), accounting for Gaussian errors on log mass, non-Gaussian errors on richness, and allowing for intrinsic scatter. In particular, since richness is determined from the difference in galaxy counts between cluster and background regions, we model the Poisson fluctuations in both the cluster and background lines of sight by fitting the observed counts in these regions, rather than the derived richness values shown in Figure~\ref{fig:Mrich}. The fitting approach follows the methodology of Andreon \& Hurn (2010), Andreon (2015, 2016), and subsequent works. Adopting uniform priors on the intercept, slope angle, and intrinsic scatter, we find:
\begin{equation}
\log n_{200,WL}  = (0.85\pm0.21) (\log \frac{M_{200}}{M_\odot} -14.788) +1.79\pm0.05 \ ,
\end{equation}
plotted in Fig.~\ref{fig:Mrich}.  The parameters exhibit negligible correlation, as shown in Fig.~\ref{fig:corner}.
As detailed in Paper II, the fit does not require modeling the weak-lensing S/N cut, and neither the modeling of the mass function a second time, as the (negligible) Eddington (1913) correction has already been accounted for during the weak-lensing mass estimate.  We verified that adding O26 back to the fitted sample would not alter the fit results, as also obvious from its location compared to the fitted scaling.

Most of the points align along a tight mass-richness scaling, and there are no outliers greater than $2\sigma$, as confirmed by modeling the scatter with a Student's $t$-distribution with 10 degrees of freedom. This distribution is robust to the presence of outliers (e.g. Andreon \& Hurn 2010) and yields identical scaling parameters. 

Formally, the intrinsic scatter around this relation is $\sigma_{\rm intr, \log n|M} = 0.06^{+0.07}_{-0.04}$ dex. However, our analysis assumes that errors are exact (see Andreon \& Hurn 2010 and Andreon 2015 for a fit of the scaling relation with uncertain errors) and the derived intrinsic scatter
is smaller than the data errors. Therefore, the data constrain only the smallness of the scatter, not its precise 68\% error, that can be as low as to have a 90\% upper
limit to be 0.05 dex, as found by Andreon (2015) on a different sample.

The tight scaling of richness with mass confirms that the outliers in the Compton $Y$-mass scaling found in Paper II are not the result of faulty mass estimates. This is because mass cannot simultaneously be incorrect for the Compton $Y$-mass scaling relation and correct for the richness-mass scaling relation.

The fit of the same data but with swapped dependent and independent variables, is
\begin{equation}
\log \frac{M_{200}}{M_\odot}  = (0.68\pm0.21) (\log n_{200,WL} - \log 50) +14.71\pm0.05
\label{myeq}
\end{equation}
As expected (e.g. Isobe et al. 1990; Andreon \& Hurn 2010), the inverse of the slope of the fitted $M_{200}|n_{200}$ 
relation, $1/0.68 = 1.5$, is not equal to the slope of the inverse $n_{200}|M_{200}$ relation, $0.85$.

\begin{figure}
\centerline{\includegraphics[width=9truecm]{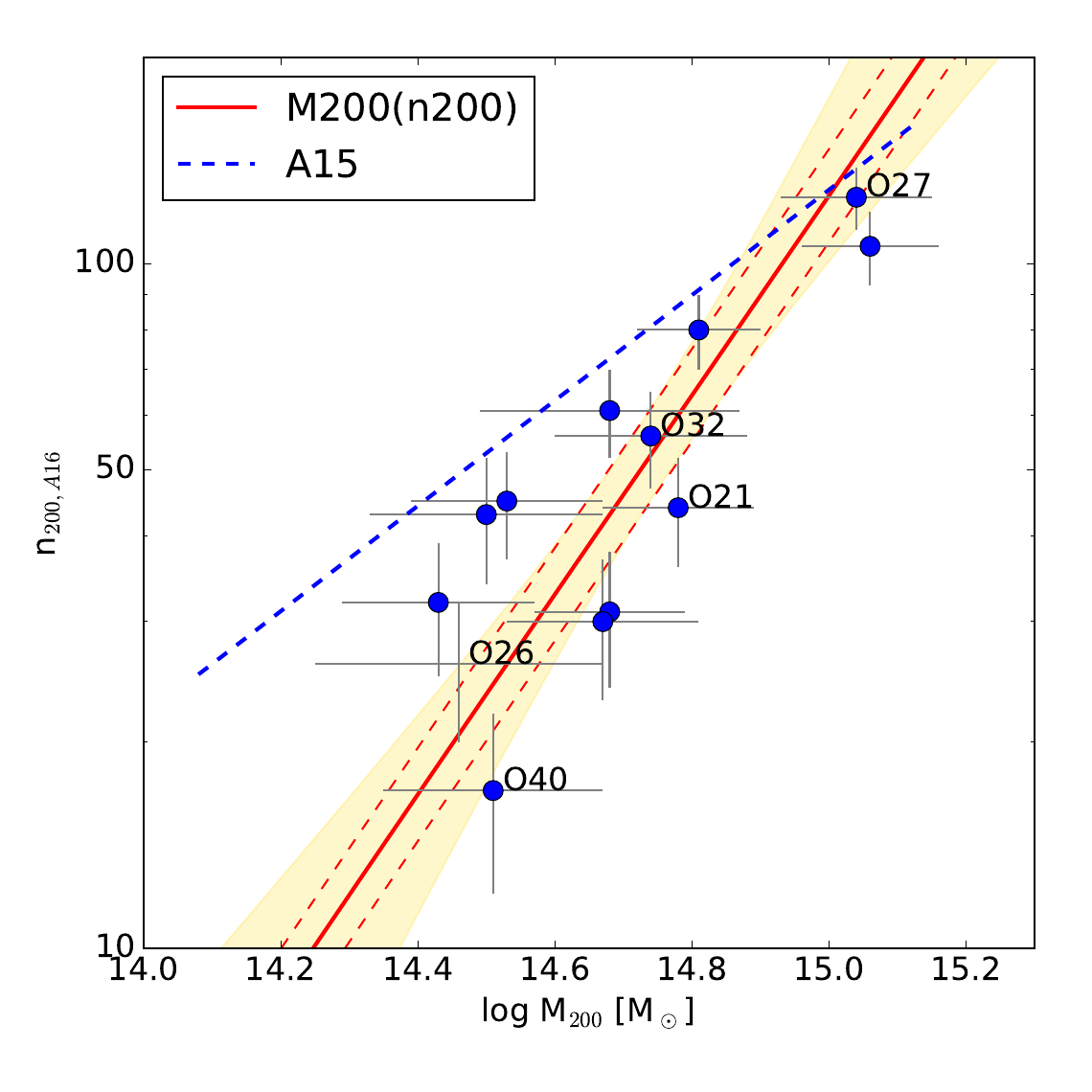}}
\caption[h]{As in Fig.~\ref{fig:Mrich}, but using richness at the galaxy-derived $r_{200}$. This richness adopts a radius uncorrelated with the weak-lensing mass estimate and a different cluster center estimate. A tight scaling is also observed here. O26 is not fitted because its possibly contaminated weak-lensing mass. Including it in the sample would not alter the scaling relation. Note that we fit mass as a function of richness in this case, but plot it with mass on the abscissa to facilitate comparison with Fig.~\ref{fig:Mrich}. The scaling relation seems to differ from the one derived using caustic masses and an X-ray selected sample (Andreon 2015, dashed line), but interpretation of the cause is premature owing
to the degeneracy between sample selection and mass bias (see text for details).}
\label{fig:MrichA16}
\end{figure}

\subsection{Results Using the richness-based aperture $r_{200,A16}$}
\label{Mn}

Figure~\ref{fig:MrichA16} presents the mass-richness relation using richness estimates, $n_{200,A16}$, derived with the galaxy-based $r_{200}$. This richness adopts an independently derived center and a radius that is uncorrelated with the weak-lensing mass estimate. Therefore, covariance induced by the use of a common radius for the measurements is absent, as there is no such common radius. This is also the relation relevant for estimating mass from richness, because when richness is used as a mass proxy, both $r_{200}$ and the center are unknown and should be determined from the galaxy counts. 

Fitting the data with the model described above, we find:
\begin{equation}
\log \frac{M_{200}}{M_\odot}  = (0.69\pm0.18) (\log n_{200,A16} - \log 50) +14.72\pm0.05
\end{equation}
with a formal intrinsic scatter of 
$\sigma_{\rm intr, \log M|n} = 0.05^{+0.06}_{-0.04}$  
dex, with the caveat that, because the errors on mass and richness are more uncertain than the intrinsic scatter estimated value, we can only conclude that the intrinsic scatter is small, but not how much close to zero. Modelling the scatter with a Student-t distribution, which is robust to the presence of outliers, returns identical scaling parameters.  We verified that adding O26 back to the fitted sample would not alter the fit results, as also obvious from its location compared to the fitted scaling. The found scaling parameters (eq. 3) are nearly identical to those derived using weak-lensing derived $r_{200}$ (eq. 2) because of the the two richnesses are nearly identical, as mentioned.

Figure~\ref{fig:MrichA16} shows a tight scaling relation, very similar to what was obtained using the weak-lensing derived $r_{200}$, indicating that our results are unaffected by radius covariance or the precise definition of the center.
The tightness of the richness-mass relation confirms and strenghten previous estimates based on caustic masses by Andreon (2016). In fact, the total scatter in the predicted mass (which corresponds to the horizontal scatter in Fig.~\ref{fig:MrichA16}) is found to be 0.11 dex, lower than the 0.16 dex measured by Andreon (2016) using caustic masses, making it even more promising as mass proxy.

The intrinsic scatter of the our richness estimator, 0.05 dex, is smaller than the intrinsic scatter 
of the Compton Y parameter for the very same sample with shared mass measurement: we fit the (log of) Compton Y 
parameter vs log mass data with a linear model with intrinsic scatter with uniform priors on intrinsic scatter, intercept, angle and log Compton Y latent values, basically the hypothesis-parsimonious model of Paper II with swapped
dependent and independent variable. We found  $\sigma_{\rm intr, \log M|Y} = 0.16^{+0.07}_{-0.06}$, confirming what found in Paper II: the relation between mass and Compton Y is quite scattered (with a larger sample in Paper III we concluded that is not a single relation at all), for our gravity-selected sample. The lower scatter of richness compared to Compton Y highlights the effectiveness of richness as
mass proxy.
As noted in Andreon (2015), the quality of a mass proxy also depends on the availability and uncertainty of the proxy observable.  Compton Y has larger errors (0.2 dex vs 0.07 dex for richness, median over the same sample) and richness has, by definition, wider availability because no SZ catalog can exist without an optical survey providing an optical identification and photometric redshifts or location of spectroscopic targets.

\begin{figure}
\centerline{\includegraphics[width=9truecm]{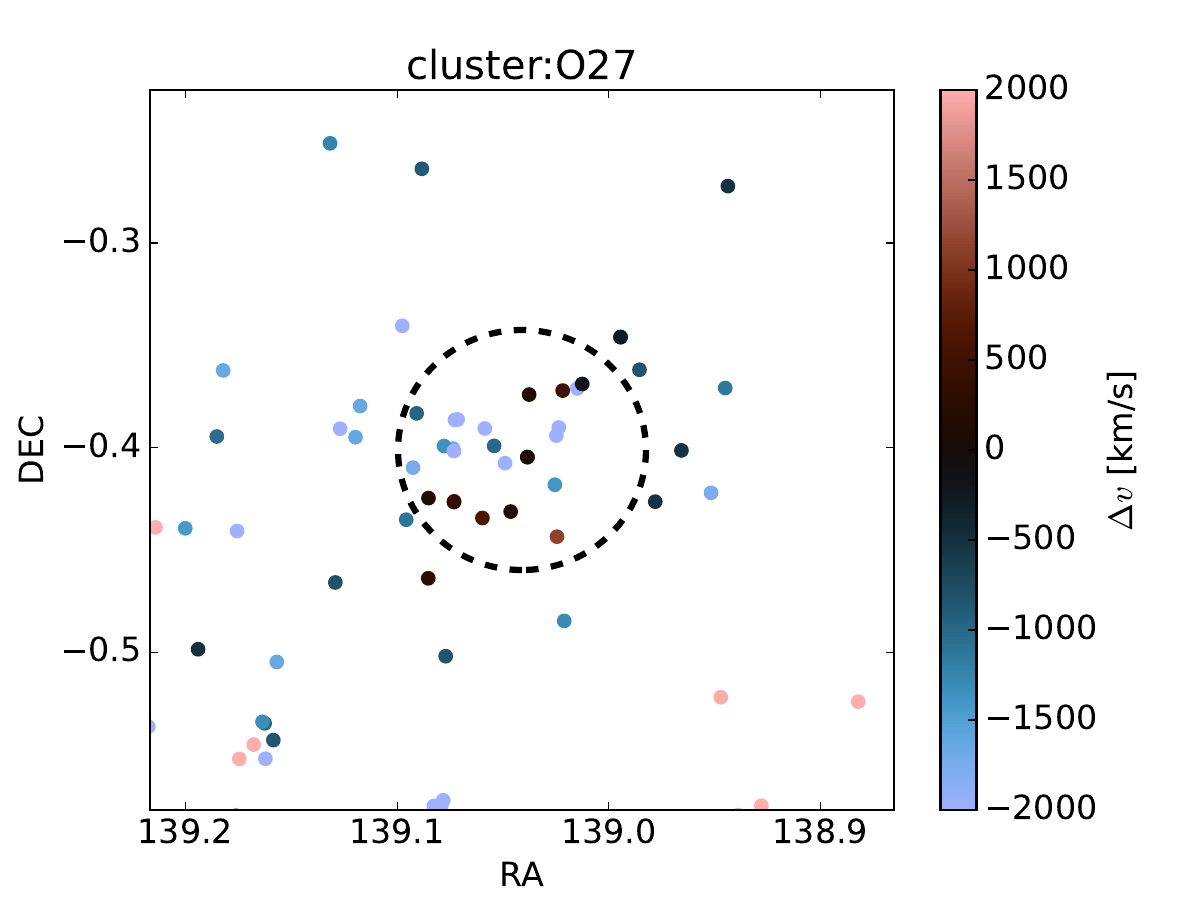}}
\caption[h]{Spatial distribution of galaxies within 3000 km/s from O27, color-coded by velocity offset. The dashed circle indicates 1 Mpc at the cluster redshift. The cluster is clearly undergoing a merger in progress.}
\label{fig:radecO27}
\end{figure}

O21, named id5 in Paper I, %,  also falls close to the Compton Y-richness scaling relation (Fig.~\ref{fig:Yrich}). It also 
falls close to the richness-mass scaling relations (Figs~\ref{fig:Mrich} and \ref{fig:MrichA16}) in apparent disagreement with what found in Paper I, where we found that its predicted mass, based on the  $n_{200,WL}$ richness, be low, by 3 sigma, from the scaling relation derived from ICM-selected samples. This is due to two factors: first, in Paper I we compare it to a different scaling relation, based on caustic masses and an X-ray
selected sample that seems
to be different, as depicted in Fig.~\ref{fig:MrichA16}. 
The comparison between the two samples is affected by a degeneracy between sample selection and mass bias. Disentangling their effects requires a sample differing in only one of these aspects, currently unavailable.
Second, the value of 
richness used in paper I underestimates the cluster richness due to an unfortunate bug: it used as total mag 
the aperture mag
that is not total for nearby clusters as O21. With the newly determined richness (i.e. after the bug correction) the cluster 
continues to be an outlier from the compared relation, with $\log M_{200,rich}/M_\odot=14.41\pm0.16$, but with lower significance (Fig.~\ref{fig:MrichA16}). The bug does not affect the other clusters in Paper I because of their higher redshifts.

O27 is also known as MACS J0916.1-0023, ACT-CL J0916.1-0024, eFEDS J091610.1-002348, and PSZ2 G231.79+31.48. 
Figure~\ref{fig:radecO27}, based on spectroscopic redshifts (DESI Collaboration, 2025), shows that it is a merger in progress, with $\Delta v \sim 2000$ km/s along the line of sight. The merging event is not expected to affect the luminosity of the quiescent galaxy population, as quiescent galaxies are nearly gas-free. Indeed, O27 has the appropriate richness for its weak-lensing mass (Fig.~\ref{fig:Mrich}), and the inverse is true as well (Fig.~\ref{fig:MrichA16}). 
Visual inspection of XMM observations and the quantitative analysis of eROSITA data by Sanders et al. (2025) confirm the disturbed morphology expected for an ongoing merger event.

\begin{figure}
\centerline{\includegraphics[width=9truecm]{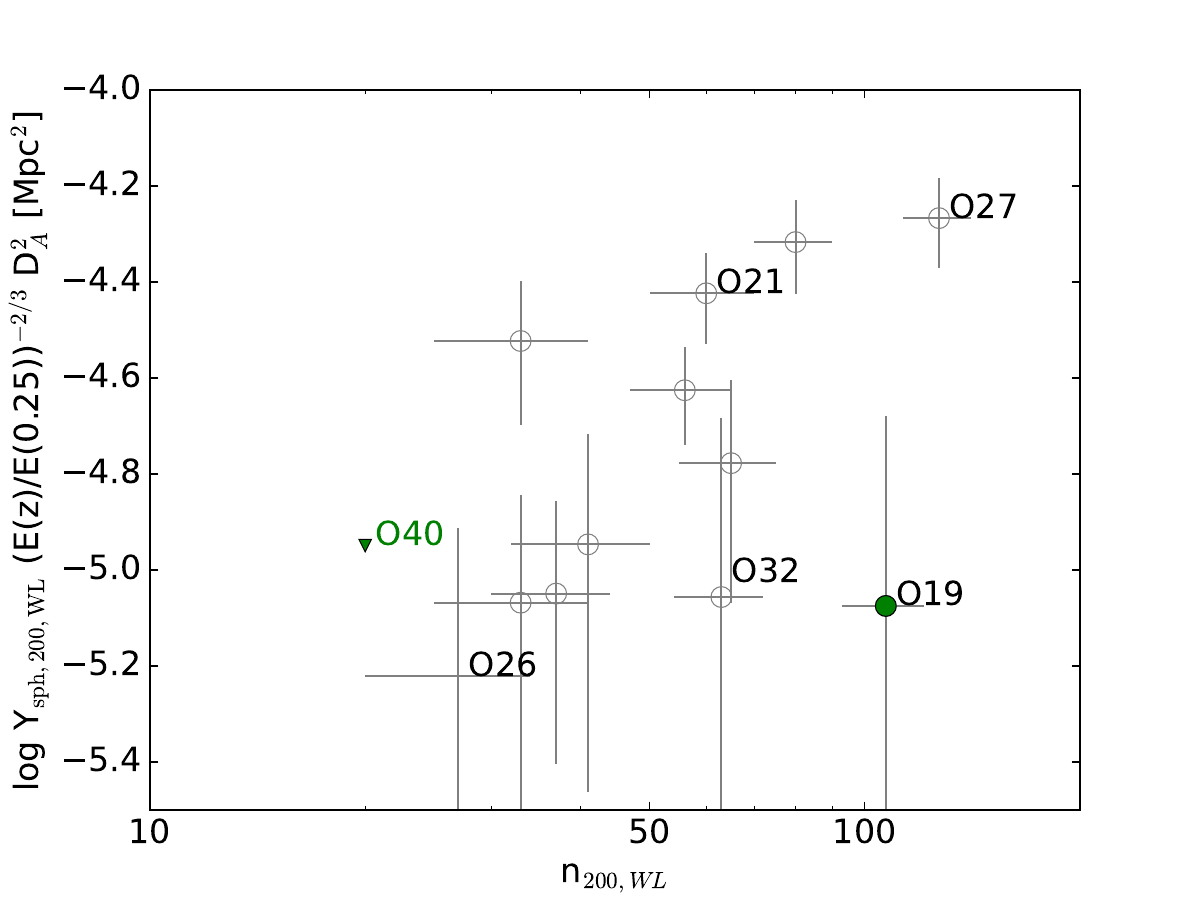}}
\caption[h]{Compton $Y$-richness scaling relation within $r_{200,WL}$. The green point and upper limit represent Compton Y-mass outliers. The data do not delineate a clear trend. Clusters discussed in the text are labeled. }
\label{fig:Yrich}
\end{figure}

\subsection{The unclear Compton $Y$-richness trend}

Once adopted the revised coordinates and redshift, O26 Compton-$Y$ parameter changed by less than $1\sigma$, from $\log Y_{{\rm sph},200} = -5.31 \pm 0.23$ to $\log Y_{{\rm sph},200} = -5.21 \pm 0.31$. We checked that using the updated parameters, or removing O26 from the fitted sample, does not alter the Compton Y-mass scaling parameters of Paper II. That analysis, as our corrent revision, uses ACT-Planck Compton $Y$ maps in Coulton et al. (2023) and accounts for the non-Gaussian nature of the errors of $\log Y_{{\rm sph},200}$, among other things.

Figure~\ref{fig:Yrich} displays the Compton $Y$ versus richness scaling relation. The data do not seem to correlate each other,
the Pearson correlation coefficient being 0.4 with a p-value of 0.18 (small p-values, say, $<0.01$, indicates the presence of a
correlation). 
As a way of comparison, the Pearson correlation coefficient between richness and mass of the very same 11 clusters
with a Compton Y detection have a Pearson correlation coefficient of 0.87 with a p-value of 0.0002, indicating that sample
size is not the only reason why the relation between Compton $Y$ and richness seems unclear.  

\begin{figure}
\centerline{\includegraphics[width=8truecm]{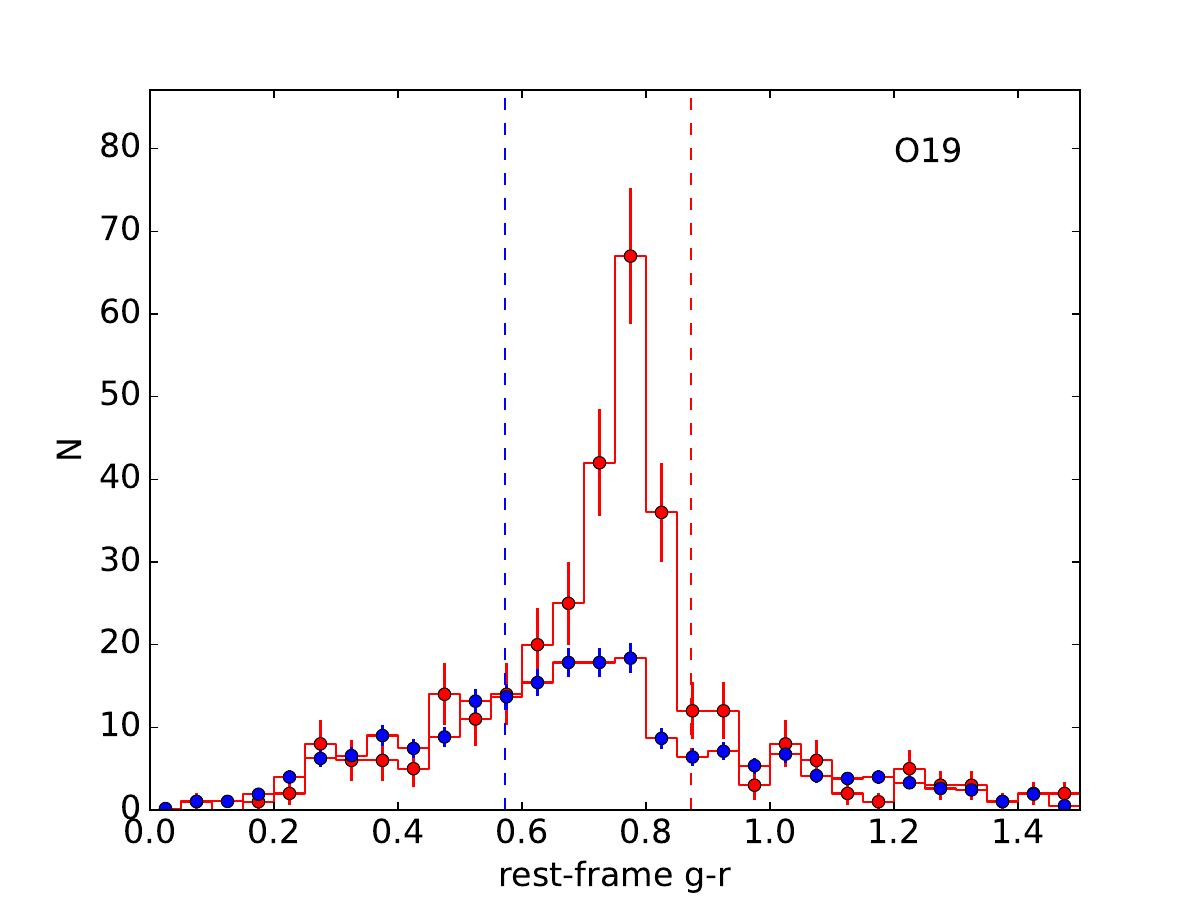}}%
\caption[h]{Color distribution of galaxies within $r_{200,\rm wl}$ (red histogram) and in a control area (normalized to the cluster solid angle, blue histogram) for O19. Only galaxies brighter than $M^e_V=-20$ mag are
considered, assuming all galaxies are at the O19 redshift.
The vertical lines indicate the expected color range of the red sequence. 
The single peak of the color distributions confirms the absence of a substantial contamination by other structures, as shown through spectroscopic data and the color-magnitude distribution in Paper II.
}
\label{fig:coldistr19}
\end{figure}

Attempts to fit the data using a range of models, including Gaussian and non-Gaussian scatter, single and mixture regressions, 
{properly handling Compton y errors, which are Gaussian on a linear scale (and therefore not Gaussian in log space)
proved highly sensitive to the choice of priors on both parameters and model forms. This sensitivity reflects the limited constraining power of the data, as also indicated by the quoted $p$-value. As a result, even the form of the Compton-$Y$ versus richness relation remains uncertain. Yet, as noted, the same 11 clusters define a remarkably tight richness-mass relation.

The lack of a correlation call for a closer scrutinity of the data.
O19, also known as the CLIO cluster, has its mass estimation and Compton Y determination scrutinized in Paper II because
it is an outlier of the Y-mass scaling. The single peak of the color distributions in the O19 line of sight (Fig.~\ref{fig:coldistr19}) confirms the absence of a substantial contamination by other structures, as shown through spectroscopic data and the color-magnitude distribution in Paper II. O19 has a $n_{200,WL}$ richness very close to the mean value expected for its mass (Fig.~\ref{fig:Mrich}), confirming
the mass estimate based on shear data. 
Discarding O19, the p-value of the Pearson correlation coefficient between richness and Compton Y decreases from 04 to 0.03, which is at most weakly suggestive of a correlation and insufficient to claim a statistically significant relationship.

Accounting for dusty star-forming galaxies, some of which reside in clusters and are spatially correlated with them, does not substantially alter the measured Compton-$Y$ values. The error-normalized residuals have an average of 0.05 (with CIB-corrected values being slightly lower) and a scatter of about 30\% of the measurement uncertainty. The CIB-corrected values were derived from the CIB-deprojected Compton-$y$ maps of Coulton et al. (2023)  with $\beta=1.7$ and $T_{CIB}=10.7$ K.

\section{Conclusions}

This paper, the third in a series, investigates the scaling relations between optical richness, weak-lensing mass, and Compton $Y$ for a sample of galaxy clusters selected purely by the effect of their gravitational potential on the shapes of background galaxies. 
Our sample comprises a complete sample of 13 gravity-selected clusters at intermediate redshifts ($0.12 \leq z_{phot} \leq 0.40$) with weak-lensing signal-to-noise ratios exceeding 7, the very same sample used in Paper II to study the Compton $Y$-mass scaling relation.
This selection method is uncommon, as most cluster samples in the literature are selected based on signals originating from cluster baryons. 

We measured cluster richness by counting red-sequence galaxies, identifying two cases of line-of-sight projections in the process, later confirmed by spectroscopic data. Both clusters, O26 and O32, are sufficiently separated in redshift that 
contamination in richness can be straighforwardly dealt because the two red sequences do not blend each other. About mass estimation, 
O32 has a foreground structure at $\Delta z \sim 0.2$, whose mass contribution is negligible.
O26 instead consists of two $\sim 3 \ 10^{14}$ M$_\odot$ clusters separated by $\sim1.5$ arcmin and $\Delta z \sim 0.2$, whose shear signal is likely affected by projection effects; we excluded this system from our reference analysis but we also consider it for testing purposes.   
Since the object exclusion is unrelated to richness, the sample without O26 continues to be gravity-selected and does not need to account for a never-applied richness selection. 
 
We find an extremely tight richness--mass relation using our red-sequence-based richness estimator, with a scatter of just $\sim0.05$ dex, smaller than %those reported for other richness estimators in the literature and smaller than 
the intrinsic scatter of Compton Y with mass for the same sample. The lower scatter highlights the effectiveness of 
richness 
compared to 
Compton $Y$. No outliers are found in the richness-mass scaling, whether or not O26 is included in the sample (with the likely contaminated mass). The O27 cluster, which is a merger in progress also obeys
the scaling, suggesting that cluster richness is not perturbed during the cluster merger.
The tightness of the richness-mass relation supports the accuracy of the weak-lensing mass measurements and reinforces the interpretation from Paper II that two clusters, O40 and O19, are genuine outliers in the Compton $Y$-mass scaling. 

In the Compton $Y$-richness plane, the data do not delineate a clear trend, as indicated by the Pearson correlation coefficient with an insignificant $p$-value, making fit results sensitive to both prions on parameters and fitted model form. The limited sample size is not the sole reason for the unclear relation between Compton $Y$ and richness, since the same sample, with identical richness values, exhibits a highly significant mass-richness correlation.

The upcoming Data Release 1 of the Euclid survey (Euclid Collaboration: Mellier et al. 2024) will enable an expansion of the gravity-selected sample by one to two orders of magnitude. Thanks to its larger size and the uniformity of richness and mass estimates, it will also allow us to break the degeneracy between sample selection and mass bias that currently limits additional investigations usig our sample.

\begin{acknowledgements}
SA acknowledges INAF grant  
``Characterizing the newly  discovered clusters of low surface  brightness" and PRIN-MIUR grant
20228B938N ``Mass and selection biases of galaxy clusters: a multi-probe approach", the latter funded by the European 
Union Next generation EU, Mission 4 Component 1  CUP C53D2300092 0006.
\end{acknowledgements}

{}

\label{lastpage}

\newpage \quad \newpage

\begin{appendix}
\section{Richness probability distribution}

\begin{table}[h]
\caption{Numerical values to derive the full richness probability distribution.}
\begin{tabular}{lrrrrrr}
\hline\hline
ID & obstot & obsbkg &  C & obstot & obsbkg &  C\\ 
& \multicolumn{3}{c}{--------- wl r$_{200}$ ----------} & \multicolumn{3}{c}{--------- A16 r$_{200}$ ---------}\\
\hline
O3 &  110 & 211 & 7.1	& 106 & 211 & 8.0    \\
O7 &  54 & 181 & 8.4   & 41 & 185 & 17.9    \\
O12 &  77 & 230 & 10.8   & 63 & 232 & 13.0  \\
O15 &  50 & 166 & 12.3   & 42 & 165 & 17.0  \\
O19 &  200 & 538 & 5.8   & 178 & 535 & 7.4  \\
O21 &  101 & 324 & 7.8   & 66 & 304 & 14.0   \\
O22 &  92 & 251 & 9.3	& 85 & 255 & 10.8    \\
O26 &  42 & 199 & 13.4   & 34 & 198 & 23.6   \\
O27 &  191 & 355 & 5.5   & 184 & 354 & 6.0   \\
O28 &  66 & 322 & 9.7	& 46 & 322 & 20.3   \\
O32 &  88 & 217 & 8.7	& 75 & 223 & 11.9   \\
O40 &  33 & 144 & 11.1   & 22 & 141 & 29.8  \\
O48 &  75 & 418 & 12.4   & 72 & 421 & 14.7  \\
\hline					     
\end{tabular}
\hfill \break 
{\bf Notes.} The full richness probability distributionw can be computed, following Andreon \& Hurn (2010), from the 
the total number of galaxies within the aperture, obstot, the number of galaxies in the control field, obsbkg, and the
ratio of their solid angles, C, for both the weak-lensing and A16 apertures. The software code is available in the JAGS language in paper above and, in PyMC, at https://github.com/PBenzoni97/Bayesian-methods-book-with-PyMC \label{tab2}
\end{table}
\end{appendix}

\end{document}